\documentclass[floats,floatfix,showpacs,amssymb,prd,twocolumn,superscriptaddress,nofootinbib,nolongbibliography,reprint]{revtex4-2}

\usepackage{amssymb,amsmath,verbatim,mathtools,needspace,etoolbox,graphicx,physics,microtype,natbib,url,hyperref,mathrsfs,bm}
\usepackage{orcidlink}

\def\mnras{{Mon.~Not.~R.~Astron.~Soc.}}

\definecolor{linkcolor}{rgb}{0.0,0.3,0.5}
\hypersetup{colorlinks=true, linkcolor=linkcolor, citecolor=linkcolor, filecolor=linkcolor, urlcolor=linkcolor}

\renewcommand{\dd}{{\,\mathrm{d}}}
\renewcommand{\det}{\mathrm{det}}
\newcommand{\pastro}{p_\mathrm{astro}}
\newcommand{\far}{\mathrm{FAR}}
\newcommand{\yr}{\,\mathrm{yr}}

\usepackage{multirow}
\usepackage{tabulary}

\usepackage[normalem]{ulem}

\newcommand{\ligo}{\affiliation{LIGO Laboratory, Massachusetts Institute of Technology, 185 Albany St, Cambridge, Massachusetts 02139, USA}}
\renewcommand{\mit}{\affiliation{Department of Physics and Kavli Institute for Astrophysics and Space Research, Massachusetts Institute of Technology,\\77 Massachusetts Ave, Cambridge, Massachusetts 02139, USA}}
\newcommand{\bham}{\affiliation{School of Physics and Astronomy and Institute for 
Gravitational Wave Astronomy, University of Birmingham,\\Birmingham B15 2TT, United Kingdom}}
\newcommand{\milan}{\affiliation{Dipartimento di Fisica ``G. Occhialini'', Universit\'a degli Studi di Milano-Bicocca, Piazza della Scienza 3, 20126 Milan, Italy}}
\newcommand{\infn}{\affiliation{INFN, Sezione di Milano-Bicocca, Piazza della Scienza 3, 20126 Milano, Italy}}

\begin{document}

\title{Calibrating signal-to-noise ratio detection thresholds using gravitational-wave catalogs}

\author{Matthew Mould\,\orcidlink{0000-0001-5460-2910}}
\email{mmould@mit.edu}
\mit \ligo

\author{Christopher J. Moore\,\orcidlink{0000-0002-2527-0213}}
\bham

\author{Davide Gerosa\,\orcidlink{0000-0002-0933-3579}}
\milan \infn \bham


\date{\today}

\begin{abstract}
Searching for gravitational-wave signals is a challenging and computationally intensive endeavor undertaken by multiple independent analysis pipelines. While detection depends only on observed noisy data, it is sometimes inconsistently defined in terms of source parameters that in reality are unknown, e.g., by placing a threshold on the optimal signal-to-noise ratio (SNR). We present a method to calibrate unphysical thresholds to search results by performing Bayesian inference on real observations using a model that simultaneously parametrizes the intrinsic network optimal SNR distribution and the effect of search sensitivity on it. We find consistency with a fourth-order power law and detection thresholds of $10.5_{-2.4}^{+2.1}$, $11.2_{-1.4}^{+1.2}$, and $9.1_{-0.5}^{+0.5}$ (medians and 90\% credible intervals) for events with false-alarm rates less than $1\,\mathrm{yr}^{-1}$ in the first, second, and third LIGO--Virgo--KAGRA observing runs, respectively. Though event selection can only be self-consistently reproduced by physical searches, employing our inferred thresholds allows approximate observation-calibrated selection criteria to be applied when efficiency is required and injection campaigns are infeasible.
\end{abstract}

\maketitle

\section{Introduction}
\label{sec: introduction}

Assembling the LIGO--Virgo--KAGRA (LVK) \cite{LIGOScientific:2014pky, VIRGO:2014yos, KAGRA:2018plz} gravitational-wave (GW) catalogs \cite{LIGOScientific:2018mvr, LIGOScientific:2020ibl, LIGOScientific:2021usb, KAGRA:2021vkt} requires searching for signals.
Classification of a signal as being of astrophysical origin typically involves information from several detectors, including the matched-filter signal-to-noise ratio (SNR) optimized over large template banks, coincidence analyses, data-quality checks, and estimates of the expected noise and astrophysical population.
Searches with the multiple independent pipelines are therefore computationally expensive \cite{Davies:2020tsx, Cannon:2020qnf, Aubin:2020goo, Klimenko:2015ypf}.

Instead of imposing a selection threshold based on data-dependent quantities such as the observed signal-to-noise ratio (SNR), false-alarm rate (FAR), or probability of astrophysical origin ($\pastro$), it is common to take ranking statistics that are far more simple to compute but that additionally depend on the true binary source parameters, such as the optimal SNR \cite{Finn:1992xs, Chernoff:1993th}.
This approach is inconsistent with the data generation process because real searches never have access to the true source properties due to parameter degeneracies and detector noise \cite{Essick:2023upv}. Nevertheless, it has
led to several approximations for rapidly computing selection effects \cite{Talbot:2020oeu, Gerosa:2020pgy, Wong:2020wvd, Mould:2022ccw, Chapman-Bird:2022tvu},
and has been employed for population inference (e.g., \cite{LIGOScientific:2018jsj, LIGOScientific:2020kqk, KAGRA:2021duu, Mould:2021xst}) and to process population-synthesis simulations (e.g., \cite{Barrett:2017fcw, COMPASTeam:2021tbl, Broekgaarden:2021efa, Dominik:2014yma, Baibhav:2019gxm}).
Doing so requires selecting a threshold value for the chosen ranking statistic that should well represent the full searches that produce observational GW catalogs.

One approach is to generate a set of fake GW signals, inject them into the search pipelines \cite{ligo_scientific_collaboration_and_virgo_2023_7890437}, filter the injections with the physical data-dependent threshold, and optimize the approximate threshold such that the resulting distribution of sources  most closely matches the filtered injections. This necessitates running the search pipelines on a large set of sources which carries a significant computational cost, which the goal of approximating selection effects was to avoid (though injection sets generated for other uses can be repurposed \cite{Essick:2023toz}). Another approach is to analytically approximate the noise distribution and its effects on physical ranking statistics such as the observed SNR \cite{Essick:2023toz}, which can be used as an efficient threshold in place of full search pipelines \cite{Fishbach:2019ckx, Farah:2023vsc}.

We present a different approach that instead makes use of observational GW catalogs to which the real searches have already been applied. Given a set of astrophysical sources that pass the pipeline threshold and their associated parameter uncertainties, we use the likelihood that the GW data were detected by the real searches but subsequently parametrize detection in terms of a ranking statistic such as the network optimal SNR. We simultaneously infer the detectable and intrinsic properties of the population of this ranking statistic, thus calibrating the unphysical detection threshold against full searches on real signals while accounting for the systematic uncertainties between the two and the statistical uncertainty in each event. Modelers who wish to apply selection effects with an unphysical model such as a threshold in the network optimal SNR can therefore use our inference, which is fully consistent with the aforementioned injection-calibration method \cite{Essick:2023toz}, to ensure the application of selection effects is representative of full searches.

In Sec.~\ref{sec: inference} we describe our Bayesian approach to simultaneous inference of the detectable and intrinsic properties of the ranking-statistic distribution, in this case the network optimal SNR. The results are presented in Sec.~\ref{sec: results} and a concluding discussion is made in Sec.~\ref{sec: discussion}.

\section{Inference}
\label{sec: inference}

In the following, we use $\pi$ to denote prior distributions, $\mathcal{L}$ to denote likelihoods (i.e., data distributions), $P$ to denote discrete distributions (e.g., for detectability which has a binary outcome), $\mathcal{P}$ for posteriors, and $p$ for mixed distributions of observed and latent variables.

\subsection{Hierarchical posterior}
\label{sec: hierarchical posterior}

We consider the joint hierarchical distribution over detectable data $\{\det_n, d_n\}_{n=1}^N$ and latent variables $\{\rho_n\}_{n=1}^N$, where the data for each GW event is labeled with a flag ``det'' denoting that it has been included in the catalog according to some criterion. Each $\rho_n$ is drawn from a common underlying population $\pi(\rho|\lambda)$ characterized by global parameters $\lambda$. While a physical model defines detectability in terms of the data alone to produce the catalog, we subsequently model detection as depending on $\rho$ and additional global parameters $\kappa$. For example, if detection is taken as a step function in $\rho$, the parameter $\kappa$ could indicate the location of the threshold.

Assuming the data from each event are independent, the GW population likelihood \cite{Mandel:2018mve, Vitale:2020aaz} is
\begin{align}
&p(\{\det_n, d_n, \rho_n\} | \kappa, \lambda)
\propto
P(\det|\kappa,\lambda)^{-N}
\notag \\
&\times
\prod_{n=1}^N
P(\det_n|\rho_n,\kappa) \mathcal{L}(d_n|\rho_n) \pi(\rho_n|\lambda)
\, ,
\label{eq: joint likelihood}
\end{align}
where $\{x_n\}$ is shorthand for $\{x_n\}_{n=1}^N$. Note that unlike a physically-motivated model for detection, i.e., $\mathcal{L}(\det,d|\rho) = P(\det|d) \mathcal{L}(d|\rho)$, detection-dependent terms $P(\det|\rho,\kappa)$ are retained above due to the assumed dependence on the parameters $\rho_n$ and $\kappa$. This is equivalent to a modification of the prior $\pi(\rho|\lambda)$, whereby regions of support are explicitly downweighted or truncated a priori \cite{Essick:2023upv}, according to Bayes' theorem:
\begin{align}
\pi ( \rho | \det, \kappa, \lambda )
=
\frac
{ P ( \det | \rho, \kappa ) }
{ P ( \det | \kappa, \lambda ) }
\pi ( \rho | \lambda )
\, .
\label{eq: pi pop det}
\end{align}
Consequently, the global detection probability
\begin{align}
P(\det|\kappa,\lambda)
=
\int P(\det|\rho,\kappa) \pi(\rho|\lambda) \dd\rho
\label{eq: pdet global}
\end{align}
does not depend on the noise model through the marginalization of the data distribution but does depend on our assumed parametrization of detection through $\kappa$. Since we will take simple functional forms for $P(\det|\rho,\kappa)$ and $\pi(\rho|\lambda)$, the integral in Eq. (\ref{eq: pdet global}) can be computed in closed form. In other words, we are using the observed data distribution, given by $\{ \det_n, d_n \}_{n=1}^N$, to infer the observable distribution of $\rho$, given by $\pi ( \rho | \det, \kappa, \lambda )$.

Before performing a population analysis with the joint likelihood above, separate parameter-estimation (PE) analyses are used to infer $\rho_n$ for each event individually under priors $\tilde{\pi}_n(\rho_n)$ that may differ event-to-event and also differ from $\pi(\rho|\lambda)$. However, using the fact that the single-event likelihood is unchanged, the individual posteriors are $\tilde{\mathcal{P}}(\rho_n|d_n) \propto \mathcal{L}(d_n|\rho_n) \tilde{\pi}_n(\rho_n)$. We can therefore write the population likelihood as
\begin{align}
&p(\{\det_n, d_n, \rho_n\} | \kappa, \lambda)
\propto
P(\det|\kappa,\lambda)^{-N}
\notag \\
&\times
\prod_{n=1}^N
P(\det_n|\rho_n,\kappa)
\frac
{ \tilde{\mathcal{P}}(\rho_n|d_n) }
{ \tilde{\pi}_n(\rho_n) }
\pi(\rho_n|\lambda)
\, .
\label{eq: joint likelihood posteriors}
\end{align}
We do not directly infer each $\rho_n$ from PE runs, which instead infer the posteriors $\tilde{\mathcal{P}}(\theta_n|d_n)$ of the source parameters $\theta_n$ under default priors $\tilde{\pi}_n(\theta_n)$. While each of the original priors for $\theta_n$ may have the same shape, they could impose different cuts in the parameter space. Each of these distributions directly implies corresponding distributions for $\rho_n$. In practice, we usually have access to the posteriors $\tilde{\mathcal{P}}(\theta_n|d_n)$ through discrete samples $\{\theta_{ni}\}_{i=1}^{N_n}$ rather than as continuous distributions, where $N_n$ is the number of posterior samples for event $n$. However, if we wish to evaluate the density in Eq. (\ref{eq: joint likelihood posteriors}), we must be able to evaluate the individual posterior densities.

Using Bayes' theorem, the joint posterior is
\begin{align}
\mathcal{P} ( \{ \rho_n \}, \kappa, \lambda | \{ \det_n, d_n \} )
\propto
p ( \{ \det_n, d_n, \rho_n \} | \kappa, \lambda ) \pi( \kappa, \lambda )
\, ,
\label{eq: joint posterior}
\end{align}
where $\pi(\kappa,\lambda) = \pi(\kappa) \pi(\lambda)$ is the prior over the population-level parameters that we take as independent. We sample the posteriors using \textsc{numpyro} \cite{Phan:2019elc}. In Sec. \ref{sec: results} we present results inferred using the joint distributions in Eq. (\ref{eq: joint likelihood posteriors}), but in Appendix \ref{app: marginal posterior} we show that they are fully consistent with the posteriors inferred using a numerically marginalized likelihood.

\subsection{Optimal signal-to-noise ratio}
\label{sec: snr}

We will take as our ranking statistic the network optimal SNR $\rho$. The optimal SNRs for each interferometer individually are given by
\begin{align}
\rho_i^2(\theta)
=
4 \int \frac{|h_i(f,\theta)|^2}{S_i(f)} \dd f
\, ,
\end{align}
where $h_i$ is the GW signal model at the detector as a function of the source parameters and frequency $f$, $S_i(f)$ is the one-sided power spectral density (PSD) of the noise, and $i$ indexes the interferometers. For a set of $I$ detectors the network optimal SNR is
\begin{align}
\rho(\theta)
=
\sqrt{ \sum_{i=1}^I \rho_i^2(\theta) }
\, .
\end{align}
We take \textsc{IMRPhenomXPHM} \cite{Pratten:2020ceb} as the waveform model and use \textsc{PyCBC} \cite{Usman:2015kfa} and \textsc{bilby} \cite{Ashton:2018jfp} to compute SNRs.

Note that the optimal SNR should also depend on the observed data via estimation of the PSD. However, we make the approximation that it is given by fixed representative noise curves for the first (O1), second (O2), and third (O3) observing runs of the LIGO Hanford\footnote{\href{https://dcc.ligo.org/LIGO-G1500622/public}{dcc.ligo.org/LIGO-G1500622/public}.}, LIGO Livingston\footnote{\href{https://dcc.ligo.org/LIGO-G1401390/public}{dcc.ligo.org/LIGO-G1401390/public}.}, and Virgo\footnote{\href{https://dcc.ligo.org/P1800374/public}{dcc.ligo.org/P1800374/public} for O2 and \href{https://dcc.ligo.org/LIGO-T2000012/public}{dcc.ligo.org/LIGO-T2000012/public} for O3.} detectors. For each event we compute the network optimal SNR with the detectors that were observing. This is simply fixing our choice of ranking statistic; one could simplify further by instead taking a single hypothetical design PSD, computing the single-detector optimal SNR, and calibrating this to the observed data, because the ranking statistic can be anything that reasonably encodes detectability.

For the joint hierarchical likelihood, we construct continuous single-event likelihood approximations. We fit each single-event posterior $\tilde{\mathcal{P}}(\rho_n|d_n)$ for the network optimal SNR as a univariate truncated normal distribution (such that $\rho_n\geq0$). Their means $\mu_n$ and variances $\sigma_n^2$ are taken as the sample means and variances of the posterior samples $\{\rho_{ni}\}_{i=1}^N$. For the single-event priors for each $\rho_n$ we take log-normal distributions and use \textsc{scipy} \cite{Virtanen:2019joe} to fit for their shape and scale parameters by minimizing the cross entropy between the true distributions implied by prior samples and the fitted approximations. This likelihood approximation is verified by comparison to a numerically marginalized likelihood in Appendix \ref{app: marginal posterior}.

\subsection{Priors}

While the individual event SNR priors $\tilde{\pi}_n(\rho_n)$ are determined by the original uninformative PE priors imposed over the source parameters, the population prior $\pi(\rho|\lambda)$ can be reasonably well-motivated astrophysically \cite{Schutz:2011tw, Chen:2014yla}. In the local universe, luminosity distances $D$ approximately follow a geometric prior $\pi(D)\propto D^2$. Since the SNR scales as $\rho\sim 1/D$ the corresponding SNR distribution scales as
\begin{align}
\pi ( \rho )
=
\pi ( D )
\left| \frac{dD}{d\rho} \right|
\sim
\rho^{-4}
\, .
\label{prho}
\end{align}
Our default model is therefore a power-law distribution in the network optimal SNR. As we are primarily interested in inferring a single threshold value in our chosen ranking statistic, our default detection model will be a Heaviside step function in $\rho$. We describe in detail the various models we consider below and summarize them in Table~\ref{tab: models}, where the model label indicates the number of its parameters (apart from the flow-based model).

\begin{table}
\centering
\begin{tabular*}{\columnwidth}{@{\extracolsep{\fill}}ccccc}
\hline
Model & Intrinsic & Detection & Parameters & Priors
\\ \hline
\multirow{2}{*}{1} & Power & \multirow{2}{*}{Heaviside} & Slope $\sigma$ & $=4$
\\
& law & & Threshold $\tau$ & $\mathcal{U}(0,20)$
\\ \hline
\multirow{2}{*}{2} & Power & \multirow{2}{*}{Heaviside} & Slope $\sigma$ & $\mathcal{U}(1,10)$
\\
& law & & Threshold $\tau$ & $\mathcal{U}(0,20)$
\\ \hline
\multirow{3}{*}{\hfill 3} & \multirow{2}{*}{Power} & \multirow{3}{*}{Sigmoid} & Slope $\sigma$ & $\mathcal{U}(1,10)$
\\
& \multirow{2}{*}{law} & & Threshold $\tau$ & $\mathcal{U}(0,20)$
\\
& & & Width $\omega$ & $\mathcal{U}(0,\tau)$
\\ \hline
\multirow{2}{*}{4} & Power & \multirow{2}{*}{Heaviside} & Slope $\sigma$ & $\mathcal{U}(1,10)$
\\
& law & & Thresholds $\tau_{1,2,3}$ & $\mathcal{U}(0,20)$
\\ \hline
F & \multicolumn{2}{c}{Normalizing flow} & Weights $w_{1,...,42}$ & $\mathcal{N}(0,1)$
\\ \hline
\end{tabular*}
\caption{Summary of the models considered for the intrinsic and detectable populations of the network optimal SNR $\rho$. Listed are the parameters of the models and corresponding priors, where $\mathcal{U}$ and $\mathcal{N}$ denote uniform and normal distributions, respectively. Apart from Model F, the model labels indicate the number of parameters.}
\label{tab: models}
\end{table}

\subsubsection{Model 1}

Our first model has a single parameter --- the detection threshold $\tau$ --- while the intrinsic distribution is a power law with slope $\sigma=4$, as above. Written explicitly, this is
\begin{align}
\pi ( \rho | \sigma=4 )
&\propto
\rho^{-\sigma}
\, , \\
P ( \det | \rho, \tau )
&=
\begin{cases}
1 \quad \mathrm{if} \quad \rho > \tau
\, , \\
0 \quad \mathrm{if} \quad \rho \leq \tau\
\, .
\end{cases}
\end{align}
We take a broad uniform prior $\pi(\tau) = \mathcal{U}(0,20)$.

\subsubsection{Model 2}

Our two-parameter model is the same as above except the slope $\sigma$ is also free to vary, for which we take a uniform prior $\pi(\sigma) = \mathcal{U}(1,10)$.

\subsubsection{Model 3}

Model 3 is used to test the assumption of a strict detection threshold by replacing it with a smooth transition function. Above the threshold we take the same power law as in Model 2, while below it the detection probability,
\begin{align}
P ( \det | \rho, \tau, \omega )
=
\begin{cases}
S ( \rho, \tau, \omega )
&\mathrm{if} \quad \tau-\omega \leq \rho \leq \tau
\, , \\
0
&\mathrm{if} \quad \rho < \tau-\omega
\, ,
\end{cases}
\end{align}
is parametrized by a sigmoid function,
\begin{align}
S ( \rho, \tau, \omega)
=
\frac{1}{2}
\left[ 1 + \sin ( \frac{\pi}{2} + \frac{\rho - \tau}{\omega} \pi ) \right]
\, ,
\label{eq: sigmoid}
\end{align}
that symmetrically ramps up to unity at $\rho=\tau$ over an interval of width $\omega$. Along with the same priors as in Model 2, we take $\pi(\omega|\tau) = \mathcal{U}(0,\tau)$.

\subsubsection{Model 4}

Given that the detection catalog contains observations from three distinct observing runs between which the detector sensitivities improved \cite{KAGRA:2013rdx}, we also consider an extension of Model 2 in which there are separate thresholds -- $\tau_1$, $\tau_2$, and $\tau_3$ -- for events in O1, O2, and O3, respectively. This model contains four parameters: one slope and three Heaviside thresholds. We take the same priors as in Model 2.

\subsubsection{Model F}

Finally, we consider a more flexible model that jointly parametrizes the detection probability and intrinsic distribution, as in Eq. (\ref{eq: pi pop det}), with a normalizing flow. In particular, we use a block neural autoregressive flow \cite{bnaf19} with a single hidden layer of ten units, composed with an exponential function to enforce positivity, as implemented in Refs. \cite{Phan:2019elc, matthew_mould_2023_7873398}. This model is deliberately overparametrized with 42 parameters (the weights of the neural network, for which we take standard normal priors) to test our assumptions on the shape of the detectable distribution for $\rho$. In particular, we can test whether a power law is a representative model for larger SNRs, the location of the peak of the distribution, and its width at lower SNRs.

\subsection{Event selection}

For consistency with the SNR computation in Sec. \ref{sec: snr}, we use the public parameter-estimation samples from the \textsc{IMRPhenomXPHM} analyses \cite{LIGOScientific:2019lzm, KAGRA:2023pio} in the combined GWTC-2.1 \cite{LIGOScientific:2021usb} and GWTC-3 \cite{KAGRA:2021vkt} catalogs.
Of these, three were observed in O1, seven in O2, and 78 in O3, while two events excluded by this requirement are GW170817 and GW190425.
GW170817 in particular is the event with the highest SNR, implying it is informative for the upper tail of the SNR distribution and could therefore impact inferred power-law slopes, but we do not expect its exclusion to impact detection thresholds which are informed by events with low SNRs.
To the best of our knowledge, the released posteriors for GW170817 do not include all parameters required to compute SNRs, meaning it could not be included in our analysis even taking a different waveform model.
Events are included in these catalogs if they are determined to be of astrophysical origin with probability $\pastro>0.5$ in at least one of the search pipelines.
Following Ref. \cite{KAGRA:2021duu}, we instead consider the events with $\far<1\yr^{-1}$ in at least one of the pipelines, which excludes some events from O3 and results in a catalog of 72 events.
We compare results using the two catalogs in Appendix \ref{app: catalog definition}.

\section{Results}
\label{sec: results}

In the following we report medians and 90\% symmetric credible intervals for inferred posteriors.

\subsection{Detection threshold}

\begin{figure}
\includegraphics[width=\columnwidth]{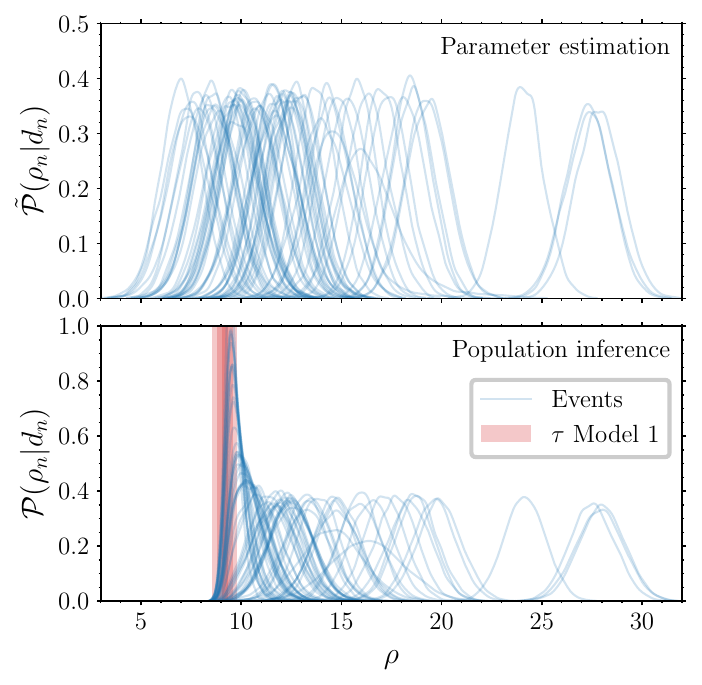}
\caption{The detection threshold $\tau$ inferred using Model 1. The shaded red bands show from darker to lighter the symmetric 50\%, 90\%, and 99\% credible intervals. The blue distributions show the network optimal SNRs $\rho_n$ as inferred by the original parameter estimation runs in the top row and by the joint population inference in the bottom row.}
\label{fig: events}
\end{figure}

In Fig.~\ref{fig: events} we display the posteriors inferred from Model 1. We find a detection threshold of $\tau=9.2_{-0.4}^{+0.4}$. Using the joint likelihood in Eq. (\ref{eq: joint likelihood posteriors}) we simultaneously infer the network optimal SNRs $\rho_n$ for each event (equivalent to population-reweighted posteriors in marginal analyses as in, e.g., Ref.~\cite{Moore:2021xhn}). The population constraint for the strict detection threshold forces the tail of lower SNR events at $\rho\approx4$ to be above threshold with $\rho\gtrsim8$.

\begin{figure}
\includegraphics[width=\columnwidth]{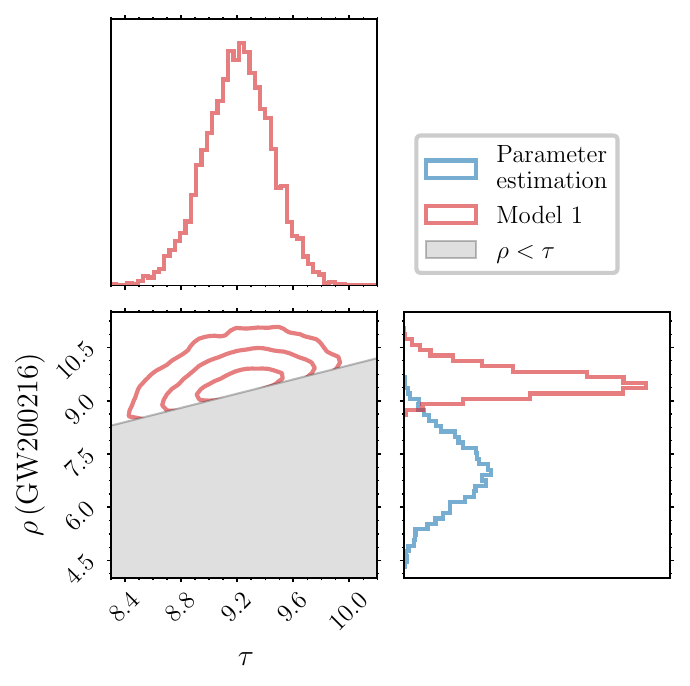}
\caption{The joint posterior distribution of the network optimal SNR $\rho$ for the event with the lowest median SNR, GW200216\_220804, and the detection threshold $\tau$ for Model 1. Red distributions show the joint population inference result, blue shows the original parameter estimation, and the gray region is excluded by the model. The panels along the diagonal show one-dimensional marginal posteriors while the lower left panel shows the two-dimensional marginal, with contours at the 50\%, 90\%, and 99\% credible levels.}
\label{fig: cut_snr}
\end{figure}

We further demonstrate this effect in Fig.~\ref{fig: cut_snr}, where we show the joint posterior for the threshold $\tau$ and the SNR for the event with the lowest median, GW200216\_220804 (shortened to GW200216 in the plot for brevity). While originally $\rho=7.0_{-1.6}^{+1.7}$ as inferred from parameter estimation samples which is largely in the region $\rho<\tau$ excluded by the assumed population model, the population inference places a constraint of $\rho=9.5_{-0.5}^{+0.7}$, the two being consistent only at the 97\% one-sided credible level.

\subsection{Astrophysical distribution}

\begin{figure}
\includegraphics[width=\columnwidth]{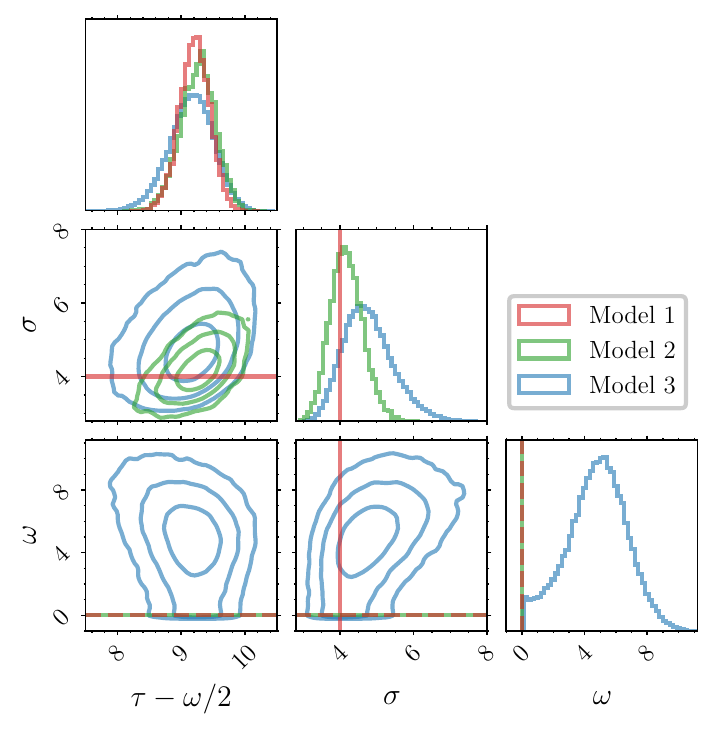}
\caption{Posterior distributions for the power-law slope $\sigma$ of the intrinsic network optimal SNR population, the centroid $\tau - \omega / 2$ of the detection threshold $\tau$, and the detection probability width $\omega$ that parametrize Models 1 (red), 2 (green) and 3 (blue). Models 1 and 2 assume $\omega = 0$, and Model 1 assumes $\sigma=4$; these constraints are displayed with vertical and horizontal lines for comparison. The diagonal panels display one-dimensional marginal posteriors and the lower left panels display two-dimensional marginals with 50\%, 90\%, and 99\% credible levels.}
\label{fig: corner}
\end{figure}

We verify the power-law slope assumed in Model 1 by letting this also be a free parameter in Model 2. Our posteriors are displayed in Fig.~\ref{fig: corner}. We find a measurement of the detection threshold only slightly broader than for Model 1 ($\tau=9.3_{-0.5}^{+0.4}$) and a slope $\sigma=4.2_{-0.7}^{+0.8}$ that is fully consistent with the astrophysical expectation of Eq. (\ref{prho}). In particular, we find that the simpler Model 1 is preferred over Model 2 with a Bayes factor of 7.7 as computed using the Savage-Dickey density ratio~\cite{01e58eb5-6ad7-3691-afed-312820f70226}.

\subsection{Detectable distribution}

We now test the detection threshold assumed by Models 1 and 2. When instead assuming that the detection probability is not a step function, we infer a width $\omega=4.9_{-3.7}^{+3.0}$, as parametrized by Model 3 in Eq. (\ref{eq: sigmoid}). This suggests a preference for nonzero widths, though we also find nonzero support for $\omega=0$ in  Fig.~\ref{fig: corner}. This broadening shifts the peak of the distribution to higher $\rho$, with $\tau=11.6_{-1.8}^{+1.4}$ for Model 3, since the nonzero density in the detectable population can be accounted for not just by the power law component but also by the sigmoid detection probability, unlike for Models 1 and 2. The centroid of the threshold remains consistent between all three models, however, with $\tau - \omega / 2 = 9.2_{-0.6}^{+0.5}$ for Model 3. This also results in a steeper but still consistent power-law slope, $\sigma=4.7_{-0.9}^{+1.3}$, since it is additionally constrained at the other end of the distribution by the absence of events with $\rho\gtrsim30$. The optimal SNR reflects only the (approximate) center of the expected distribution of observed SNRs. But beyond this, the catalog is constructed using a different ranking statistic, in this case the FAR, which does not directly translate into a strict threshold in other quantities, such as the network optimal SNR. The $\omega=0$ slice of the Model 3 posterior remains consistent with Models 1 and 2, as seen in Fig. \ref{fig: corner}.

\begin{figure}
\includegraphics[width=\columnwidth]{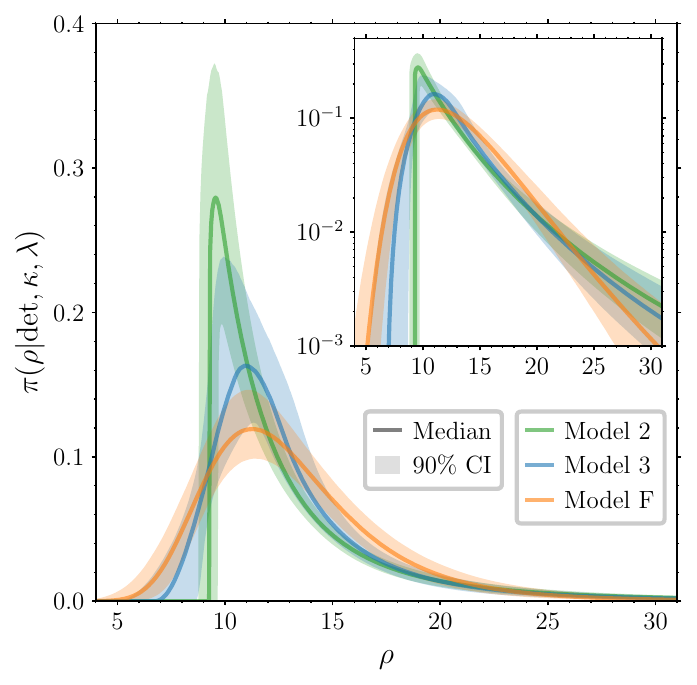}
\caption{Detectable distributions of the network optimal SNR $\rho$ and associated posterior uncertainty inferred using Models 2 (green), 3 (blue), and F (orange). Solid lines indicate the posterior median and shaded regions the symmetric 90\% credible intervals.}
\label{fig: ppds}
\end{figure}

The resulting posterior population distribution, computed by evaluating the prior models $\pi(\rho|\det,\kappa,\lambda)$ at draws from the posterior $\mathcal{P}(\kappa,\lambda|\{d\})$, is displayed in Fig.~\ref{fig: ppds}. Compared to Model 2, the distribution inferred using Model 3 has larger uncertainties, suggesting the parametrization of the former may be overly restrictive. They are consistent at larger SNRs where they assume the same parametrization, but below the detection threshold inferred by Model~2 become inconsistent beyond the 90\% uncertainty.

We verify the assumptions of Model 3 by placing a more flexible prior over the detectable population of $\rho$ with Model F, whose inferred distribution is also displayed in Fig.~\ref{fig: ppds} for comparison. The results for Models 3 and F are broadly consistent, peaking at the same location ($11.3_{-0.8}^{+0.8}$ for Model F, cf. above) and staying consistent below and above it within the 90\% posterior uncertainty, though the latter prefers a broader distribution overall. With Model F we find a slower rise with a width $5.8_{-1.3}^{+1.5}$ below the peak of the distribution.
The overall uncertainty in Model F is lower than that of Model 3 since, due to its over parametrization and resulting sparsity \cite{10.5555/3546258.3546499}, many weight posteriors are prior driven. However, we use this model parametrized by a neural network only to broadly check that the functional forms assumed for the previous models are reasonable. Choosing a smaller network does not change the resulting posterior population distribution, but changing the shape of the priors to, e.g., uniform, result in slightly larger uncertainties around the peak of the distribution.

\subsection{Differences between observing runs}

\begin{figure}
\includegraphics[width=\columnwidth]{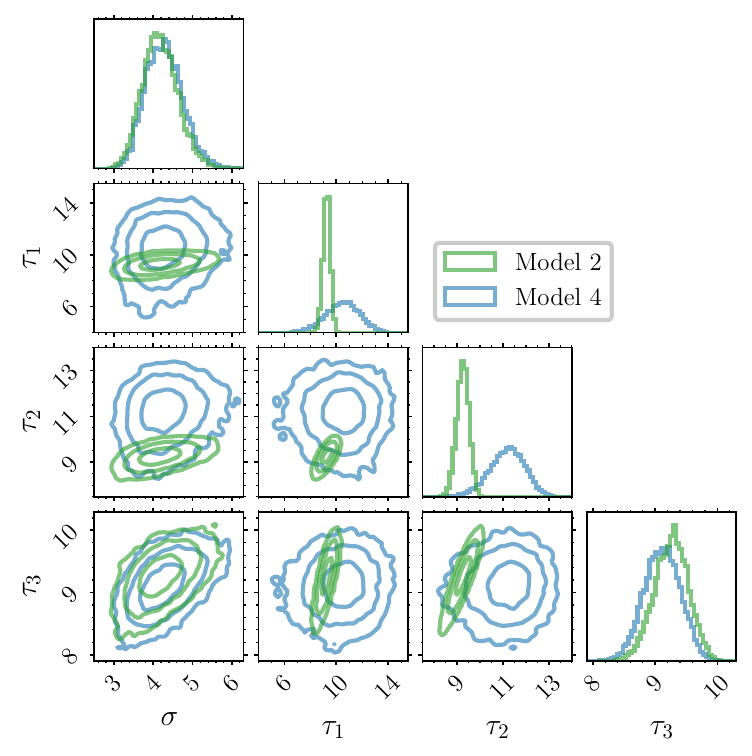}
\caption{Posteriors of the parameters for Models 2 (green) and 4 (blue). Both are characterized by a power law with slope $\sigma$ and a strict detection threshold $\tau$, shared between all three observing runs for Model 2 but separated into O1, O2, and O3 for Model 4 ($\tau_1$, $\tau_2$, and $\tau_3$, respectively). Diagonal panels show each one-dimensional posterior while lower left panels show two-dimensional marginals
with 50\%, 90\%, and 99\% credible regions.}
\label{fig: 2v4}
\end{figure}

Finally, we consider another extension to Model 2 that still assumes a strict detection threshold, but takes different values for the three LVK observing runs (Model 4). The posterior for Models 2 and 4 are compared in Fig.~\ref{fig: 2v4}. Assuming a shared power-law slope $\sigma$ leaves its measurement unchanged, with $\sigma=4.3_{-0.7}^{+0.8}$. The detection threshold for O3, $\tau_3=9.1_{-0.5}^{+0.5}$, is fully consistent with the shared thresholds in Model 1 and 2, implying the inferences for latter are primarily driven by the O3 events that make up the majority of the catalog. The thresholds for O1 and O2 are more uncertain, however, due to the lower number of events (3 and 7, respectively), with $\tau_1=10.5_{-2.4}^{+2.1}$ and $\tau_2=11.2_{-1.4}^{+1.2}$. These thresholds are consistent with each other and, while their posterior constraints do overlap with that for O3, they are typically larger. This means a signal passing the FAR cut in O3 could be included in the catalog with a lower equivalent network optimal SNR than it would in O1 or O2, given the assumed representative PSDs.

\section{Discussion}
\label{sec: discussion}

In both inference and modeling settings, GW detectability is very often unphysically thresholded using the (network) optimal SNR, a quantity that depends on the data only through estimation of the detector noise properties and moreover on the true binary source parameters. This is in contrast to real searches that depend only on the observed GW data. By simultaneously parametrizing the detection probability and intrinsic distribution of a selected ranking statistic such as the network optimal SNR, we showed that corresponding detection thresholds can be calibrated against observational catalogs that were assembled using real search pipelines. In other words, we inferred the threshold of an unphysical model that best reproduces the physical model.

However, we emphasize that the only way to fully reproduce search results is to process simulated sources with software injections and there is no guarantee that, e.g., an optimal SNR threshold will filter sources across the parameter space in the same way. Such thresholds make strong assumptions on the dependence of detectability on the source parameters and reference PSD through the SNR calculation that searches do not. In reality, detection also depends on other parameters that this method does not account for, such as the background rate which is also inhomogenous across the binary parameter space (e.g., more massive binaries produce shorter signals which are more easily mistaken for noise transients, and vice versa). Because we do not know the true source properties of search triggers, the only way to benchmark unphysical selection criteria is to construct the distribution of physical ranking statistics given the unphysical one, which would anyway require performing software injections in many noise realizations. The primary purpose of this work, then, is not to replace injection campaigns where accurate reproduction of search pipelines is necessary (e.g., population inference), but to infer appropriate thresholds for unphysical models when efficiency is paramount (e.g., processing large population synthesis simulations into mock catalogs). Our parametrization also provides a convenient functional form for the detection probability tailored to current observations.

Other approaches that model detection with some approximation for the noise properties allow selection effects to be computed while depending only on observables as opposed to the true source properties \cite{Fishbach:2019ckx, Farah:2023vsc, Essick:2023upv}. If one were to construct observational catalogs in this way, by thresholding on, e.g., the observed SNR, it is trivial to perform self-consistent downstream analyses, such as population inference.
However, the assumptions that produce (semi) analytic detection estimates, though reasonable, are not exact.
Moreover, catalogs are more commonly assembled using statistics that take account of information beyond just the SNR, such as the FAR \cite{KAGRA:2021duu} or $\pastro$ \cite{LIGOScientific:2021usb, KAGRA:2021vkt}.
When applied to such catalogs, the above approach therefore still requires the choice of an approximating threshold, because the ranking statistics used to assemble the catalog and then to estimate detection biases for, e.g., population inference or simulated reconstruction of observed events, are different.
The method we present is general and can be used to infer catalog-calibrated detection thresholds no matter the chosen approximating model.

We find a marked agreement between our inferred O3-only network optimal SNR threshold with that found by optimization against injections \cite{Essick:2023toz} --- $9.1_{-0.5}^{+0.5}$ and 9.2,
respectively (see their Fig. 6) ---
despite systematic differences between the two. Firstly, our catalog is composed of real signals that are compared to a waveform model, whereas injections are both generated and recovered using waveform models. Secondly, while both methods took fixed PSDs to calculate SNRs, Ref. \cite{Essick:2023toz} also uses those to draw noise realizations for injections, whereas there is some variability in real observations. Thirdly, the observational catalog is contaminated at some level by noise transients \cite{KAGRA:2021duu}, which is not true of injection campaigns. Our posterior measurements are therefore conditioned on this systematic waveform and noise mismodeling (though we expect the systematic uncertainty to be subdominant with respect to the statistical uncertainty), meaning downstream use of the same ranking statistic and thresholds are self consistent.

We encoded the full vector of binary source properties into the population distribution of a single representative variable, here taken to be the network optimal SNR. Even if the intrinsic population is uncorrelated between source parameters, this is not true of the observed population. A full joint inference of all binary source parameters along with global parameters that characterize the detectable and intrinsic populations will be an interesting approach for future work.

\section*{Acknowledgments}

We thank Salvatore Vitale, Jack Heinzel, and Maya Fishbach for helpful comments.
M.M. is supported by LIGO Laboratory through the National Science Foundation award PHY-1764464.
C.J.M. acknowledges the support of the UK Space Agency Grant No. ST/V002813/1.
D.G. is supported by 
ERC Starting Grant No.~945155--GWmining, 
Cariplo Foundation Grant No.~2021-0555, 
MUR PRIN Grant No.~2022-Z9X4XS, 
MSCA Fellowship No.~101064542--StochRewind, 
MSCA Fellowship No.~101149270--ProtoBH,
Leverhulme Trust Grant No.~RPG-2019-350,
and the ICSC National Research Centre funded by NextGenerationEU.   
Computational work was performed at CINECA with allocations through INFN and Bicocca.

This research has made use of data or software obtained from the Gravitational Wave Open Science Center (gwosc.org), a service of the LIGO Scientific Collaboration, the Virgo Collaboration, and KAGRA. This material is based upon work supported by NSF's LIGO Laboratory which is a major facility fully funded by the National Science Foundation, as well as the Science and Technology Facilities Council (STFC) of the United Kingdom, the Max-Planck-Society (MPS), and the State of Niedersachsen/Germany for support of the construction of Advanced LIGO and construction and operation of the GEO600 detector. Additional support for Advanced LIGO was provided by the Australian Research Council. Virgo is funded, through the European Gravitational Observatory (EGO), by the French Centre National de Recherche Scientifique (CNRS), the Italian Istituto Nazionale di Fisica Nucleare (INFN) and the Dutch Nikhef, with contributions by institutions from Belgium, Germany, Greece, Hungary, Ireland, Japan, Monaco, Poland, Portugal, Spain. KAGRA is supported by Ministry of Education, Culture, Sports, Science and Technology (MEXT), Japan Society for the Promotion of Science (JSPS) in Japan; National Research Foundation (NRF) and Ministry of Science and ICT (MSIT) in Korea; Academia Sinica (AS) and National Science and Technology Council (NSTC) in Taiwan.

\appendix

\section{Marginal posterior}
\label{app: marginal posterior}

Instead of simultaneously inferring $\kappa$, $\lambda$, and $\{\rho_n\}$, if we are uninterested in the event-level parameters we can marginalize the likelihood over each to obtain
\begin{align}
&\mathcal{L} ( \{ \det_n, d_n \} | \kappa, \lambda )
\propto
P(\det|\kappa,\lambda)^{-N}
\notag \\
&\times
\prod_{n=1}^N
\int P(\det_n|\rho_n,\kappa) \mathcal{L}(d_n|\rho_n) \pi(\rho_n|\lambda) \dd\rho_n 
\, .
\label{eq: marginal likelihood}
\end{align}
Rewriting the individual-event likelihoods in terms of the original parameter-estimation posteriors, we have
\begin{align}
&\mathcal{L} ( \{ \det_n , d_n \} | \kappa, \lambda )
=
P(\det|\kappa,\lambda)^{-N}
\notag \\
&\times
\prod_{n=1}^N
\frac{1}{N_n} \sum_{i=1}^{N_n}
P ( \det_n | \rho_{ni}, \kappa )
\frac
{ \pi(\rho_{ni} | \lambda ) }
{ \tilde{\pi}_n ( \rho_{ni} ) }
\, ,
\label{eq: marginal likelihood numeric}
\end{align}
where each integral is approximated as a Monte Carlo average over the posterior samples $\{ \rho_{ni} \}_{i=1}^{N_n} \sim \tilde{\mathcal{P}} ( \rho_n | d_n )$. The advantage of this approach is that we do not need to use fitted posteriors for the individual events as above, instead relying directly on the posterior samples. We do still require the original prior densities, but since they do not depend on $\kappa$ or $\lambda$ the prior densities evaluated at the posterior samples, $\{ \{ \tilde{\pi}_n(\rho_{ni}) \}_{i=1}^{N_n} \}_{n=1}^N$, can be computed once and stored ahead of time.

\begin{figure}
\includegraphics[width=\columnwidth]{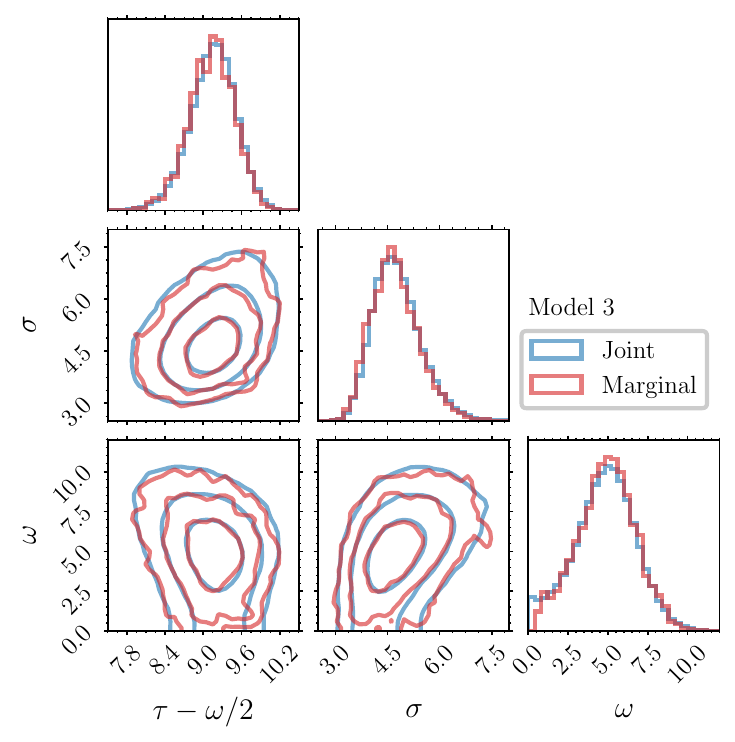}
\caption{Posterior distributions for the parameters of Model 3 as inferred in a fully hierarchical analysis (blue) and a numerically marginalized one (red). Diagonal panels contain histograms for each parameter individually while the others contain two-dimensional posteriors with contours at the 50\%, 90\%, and 99\% credible levels.}
\label{fig: joint_marginal}
\end{figure}

As an example, we show the Model 3 posteriors resulting from the full hierarchical distribution of Eq. (\ref{eq: joint likelihood posteriors}) and the marginal likelihood of Eq. (\ref{eq: marginal likelihood numeric}) in Fig.~\ref{fig: joint_marginal}. The two are almost identical, implying our assumed single-event truncated-normal posteriors are valid. The only slight discrepancy between them is that the posterior for the detection-threshold width $\omega$ inferred with the joint likelihood has nonzero support at $\omega=0$, whereas that for the marginal likelihood does not. This is likely because the truncated normal posteriors have support for $\rho_n$ across the positive real line whereas the parameter-estimation results are subject to finite-sampling effects.

\section{Catalog definition}
\label{app: catalog definition}

\begin{figure}
\includegraphics[width=\columnwidth]{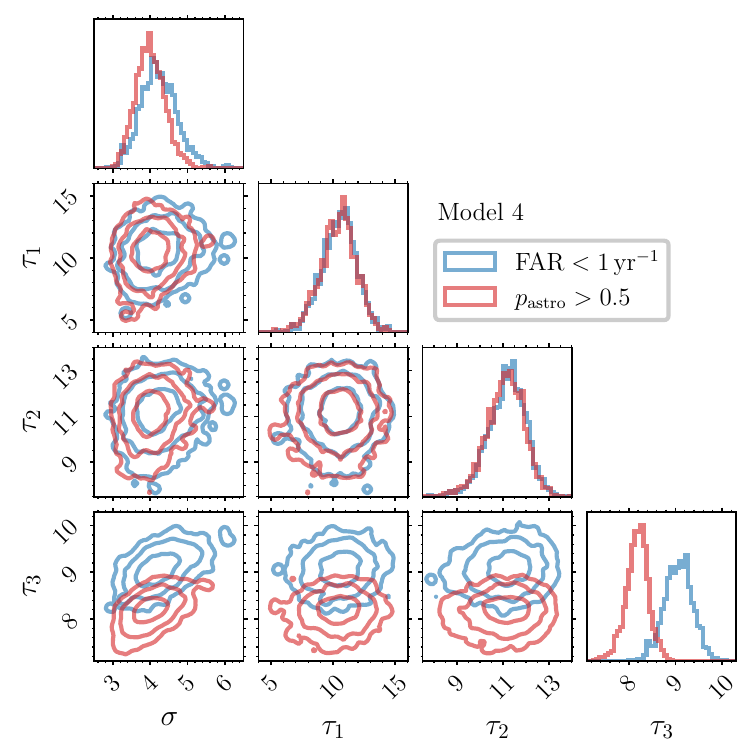}
\caption{Posterior distributions for the parameters of Model 4 inferred using catalogs where events are selected based on a threshold of $\far<1\yr^{-1}$ (blue) or $\pastro>0.5$ (red). One- and two-dimensional marginal posteriors are given in the diagonal and lower-left panels, respectively, with the latter containing containing credible regions at the 50\%, 90\%, and 99\% levels.}
\label{fig: far_pastro}
\end{figure}

The results in Sec. \ref{sec: results} are all produced using a catalog where events are selected if they satisfy $\far<1\yr^{-1}$, as in Ref. \cite{KAGRA:2021duu}. In other words, we are inferring the parameters $\kappa$ and $\lambda$ of the detectable distribution of network optimal SNRs $\rho$ that best reproduces this FAR threshold. Another common catalog threshold is for events to be more likely of astrophysical origin than otherwise, i.e., $\pastro>0.5$ \cite{KAGRA:2021vkt}. We assess the impact of these two choices in Fig.~\ref{fig: far_pastro} by running the inference with Model 4 on each catalog. In the catalog defined by $\pastro>0.5$ there are several events whose network optimal SNR posteriors clearly do not satisfy the assumed truncated-normal approximation employed in Section \ref{sec: snr}. We therefore use the marginal likelihood from Eq. (\ref{eq: marginal likelihood numeric}) in which the finite set of parameter-estimation samples are used directly, thus avoiding approximating the posterior distributions.

The only events that do not make the $\far<1\yr^{-1}$ threshold compared to the larger $\pastro>0.5$ catalog are from O3. Therefore, the parameters $\tau_1$ and $\tau_2$ that define the detection thresholds for O1 and O2, respectively, are unchanged with respect to this choice of catalog. Due to the inclusion of more low-SNR events from O3, the corresponding threshold $\tau_3$ is lowered from $\tau_3=9.1_{-0.5}^{+0.5}$ to $\tau_3=8.2_{-0.4}^{+0.4}$. The power-law slope is also mostly influenced by the O3 events in the catalog due to the larger number of them, and becomes $\sigma=4.0_{-0.6}^{+0.7}$ for the $\pastro>0.5$ catalog, compared to $\sigma=4.2_{-0.7}^{+0.9}$ for $\far<1\yr^{-1}$. In both cases the uncertainties are slightly reduced due to the addition of 18 events.

\bibliography{draft}

\end{document}